\documentclass[a4paper,fleqn]{cas-sc}

\usepackage{ensembleV}
\usepackage[authoryear,longnamesfirst]{natbib}

\def\wid{0.75\linewidth}

\newcommand\blfootnote[1]{%
	\begingroup
	\renewcommand\thefootnote{}\footnote{#1}%
	\addtocounter{footnote}{-1}%
	\endgroup
}

\begin{document}
\let\WriteBookmarks\relax
\def\floatpagepagefraction{1}
\def\textpagefraction{.001}

\shorttitle{Solar Wind Speed Estimate with Machine Learning Ensemble Models for LISA}    

\shortauthors{F. Sabbatini, C. Grimani}  

\title[mode = title]{Solar Wind Speed Estimate with Machine Learning Ensemble Models for LISA}

\author{Federico Sabbatini}[orcid=0000-0002-0532-6777]\cormark[1]\ead{f.sabbatini1@campus.uniurb.it; (+39)0722-303383}

\author{Catia Grimani}[orcid=0000-0002-5467-6386]\ead{catia.grimani@uniurb.it}

\address{Department of Pure and Applied Sciences, University of Urbino Carlo Bo, Via S. Chiara, 27, Urbino, 61029, Italy}
\address{Section in Florence, INFN, Via B. Rossi, Sesto Fiorentino, 50019, Florence, Italy}

\cortext[1]{Corresponding author.}

\begin{abstract}
	In this work we study the potentialities of machine learning models in reconstructing the solar wind speed observations gathered in the first Lagrangian point by the ACE satellite in 2016--2017 using as input data galactic cosmic-ray flux variations measured with particle detectors hosted onboard the LISA Pathfinder mission also orbiting around L1 during the same years.
	We show that ensemble models composed of heterogeneous weak regressors are able to outperform weak regressors in terms of predictive accuracy.
	Machine learning and other powerful predictive algorithms open a window on the possibility of substituting dedicated instrumentation with software models acting as surrogates for diagnostics of space missions such as LISA and space weather science.
\end{abstract}


\maketitle

\section{Introduction}

\blfootnote{List of abbreviations.
	ANN: artificial neural network.
	BR: Bartels rotation.
	ESA: European Space Agency. 
	GCR: galactic cosmic ray.
	HSS: high-speed streams.
	ICME: interplanetary counterpart of coronal mass ejection.
	IMF: interplanetary magnetic field.
	$k$-NN: $k$-nearest neighbors.
	LISA: Laser Interferometer Space Antenna.
	LR: linear regressor.
	LPF: LISA Pathfinder.
	MAE: mean absolute error.
	ML: machine learning.
	MLP: multilayer perceptron.
	RBF: radial basis function.
	RF: random forest.
	SVM: support vector machine.
}
Recent years have seen an exponential increase in the application of machine learning (ML) techniques to several fields, including physics and astrophysics~\citep{Nguyen2019AutomaticDO,Zhou2021ARO,Reiss2021MachineLF,Rdisser2022AutomaticDO}.
ML-based tools provide predictions more accurate than other models due to their generalisation properties.
Furthermore, the current availability of computational power makes feasible building complex and prediction-effective predictors.
These are the main reasons behind the heavy application of ML algorithms, even though they usually require a huge amount of training data to be provided.
This need of large training data sets may be challenging in some scenarios, but it does not constitute a limitation when ML predictors rely on observations gathered on beam experiments in high-energy physics and by long-lasting space missions, for instance.

Predictive tools in general, and ML models in particular, are precious resources for space missions to achieve several goals, as time series missing data filling and pattern recognition~\citep{Villani22}.
Even in the case that some data are not available at all, ML predictors may allow to provide these data.
The future European Space Agency (ESA) Laser Interferometer Space Antenna~\citep[LISA;][]{lisa} mission, devoted to the detection of low-frequency gravitational waves in space, will host magnetometers and radiation monitors for diagnostics~\citep{cesarini}.
However, no instruments dedicated to the measure of solar wind speed will be placed onboard LISA~\citep{apj1,apj2,Villani22,cesarini}.
We show here that a predictive ML model may allow us to estimate the solar wind speed variations for LISA by using, as input data, galactic cosmic-ray (GCR) flux observations that appear modulated at the transit of high-speed solar wind streams.
We have developed a ML predictive model for LISA based on the LISA Pathfinder \citep[LPF;][]{Armano2009,antonucci2011,lisapf2,armano2016,armano18,bridge} data gathered between February, 2016 and July, 2017 around the L1 Lagrangian point.
The outcomes of the model are compared to observations gathered contemporaneously by the NASA mission ACE~\citep{stone2013} dedicated to interplanetary medium parameter monitoring, also orbiting around L1 during the same period of time of LPF.
It is worthwhile to point out that ACE observations can be used also for LPF because the distance between the two spacecraft was always smaller than the solar wind correlation length~\citep{wicks}.
In this paper we show that GCR flux short-term recurrent variations allow to reconstruct the solar wind speed trend during the time LPF remained into orbit with an ML ensemble predictor.
The same ML ensemble predictor will be used for solar wind speed estimates for LISA after the mission launch scheduled in 2037.
Accordingly, in \Cref{sec:rel} background notions on LPF and LISA missions are provided, in \Cref{sec:contr} the ensemble model is described in detail, from the data pre-processing through the model tuning and testing, and in \Cref{sec:conc} conclusions are drawn.

\section{Background and Motivations}\label{sec:rel}

\subsection{The LISA and LISA Pathfinder missions}

The LISA mission will be the first interferometer for the detection of low-frequency gravitational waves in space.
It will consist of a constellation of 3 spacecraft placed in a triangular formation of 2.5 million km to the side, corresponding to the arm of the interferometer.
Each spacecraft will host particle detectors, platform magnetometers and temperature sensors for the diagnostics of spurious forces acting on test masses playing the role of the interferometer mirrors~\citep{cesarini}.
The LISA proof masses are 2~kg free-falling cubes of 4.6 cm side made of a gold-platinum alloy.

The LISA precursor mission, LISA Pathfinder, was aimed to test the technology to be hosted onboard LISA.
The LPF spacecraft was launched on December, 2015 from the Kourou base in French Guiana.
It followed a quasi-elliptic orbit around the first Lagrangian point at 1.5 million km distance from Earth in the Earth-Sun direction.
A particle detector optimised to monitor GCRs and solar particles with energies above 70 MeV/n was placed onboard the spacecraft~\citep{can11}.
The instrument consisted of 2 silicon wafers of 1.4 x 1.05 cm$^2$ area and thickness equal to 300 $\mu$m placed in a telescopic arrangement at 2 cm distance inside a shielding copper box of 6.4 mm thickness, to stop particles with energies $<$~70~MeV~n$^{-1}$.
The LPF radiation monitor observations of the GCR integral flux were gathered at the rate of 15 s and binned hourly presented a statistical uncertainty of 1\%.

\subsection{Galactic cosmic-ray flux and solar wind speed}

The observed cosmic-ray flux in the inner heliosphere consists almost entirely of protons (90\%) and helium nuclei (8\%) of galactic origin and it is observed to undergo long- and short-term modulations~\citep{papini96}.
Long-term modulations are associated with the solar cycle and polarity, with periodicities of 11 and 22 years, respectively~\citep{grim07}.
GCR short-term variations are those characterised by durations shorter than 27 days, that is the average solar rotation period observed from Earth.
These GCR depressions are associated with the passage of interplanetary large-scale structures and are observed when the solar wind speed increases above 400~km~s$^{-1}$ and/or the interplanetary magnetic field (IMF) intensity shows values larger than 10~nT.
Such threshold values of the mentioned interplanetary parameters are crossed at the passage of solar wind high-speed streams (HSS) and interplanetary counterparts of coronal mass ejections~\citep[ICMEs; see][and references therein]{apj1}.
The correlation between IMF, solar wind speed increases and GCR flux depressions suggests to adopt ML models or other techniques to reconstruct the trend of one parameter on the basis of the variations of the others.
For instance, \cite{Villani22} considered different methods to reproduce GCR flux observations gathered onboard LPF starting from contemporaneous observations of the solar wind speed and/or the IMF intensity gathered by the ACE experiment.
The final aim of this work is to control the energy-dependent GCR flux variations to correctly estimate the charging and the charging noise of the LISA test masses~\citep{bridge}.

\subsection{ML predictors for the solar wind}

Several works are found in the literature involving ML techniques to obtain predictions of solar wind parameters, or to reconstruct the observed solar wind data.
Existing works focus in particular on solar wind parameter classification~\citep{Camporeale2017,HEIDRICHMEISNER2018397,Roberts2020,Li2020} with few exceptions for regression tasks~\citep{Upendran20,Kataoka21}.
More in detail, \cite{Kataoka21} exploit an echo state network to reproduce the solar wind corresponding to extreme magnetic storms by using as input data geomagnetic activity indices.
A different approach is presented in~\cite{Upendran20}, where a deep neural network has been applied to solar corona ultraviolet images for the prediction of solar wind speed observations carried out in the first Lagrangian point.

The work presented here differs from both mentioned approaches, being based on an ML ensemble regressor allowing us to reconstruct the solar wind speed observed in the first Lagrangian point by using as input data the GCR flux variation measurements.

\section{An ML Ensemble Model for Solar Wind Speed Estimate}\label{sec:contr}

\begin{table}
	\begin{tabular}{c@{\hskip 0.5in}cccc}
		\toprule
		Name & Moving average & Granularity & GCR observations & GCR time window length \\
		& (h) & (h) & (n) & (d) \\
		\midrule
		DS1 & 3 & 3 & 24 & 3 \\
		DS2 & 3 & 3 & 48 & 6 \\
		DS3 & 5 & 5 & 24 & 5 \\
		DS4 & 5 & 5 & 48 & 10 \\
		\bottomrule
	\end{tabular}
	\caption{Data sets used to train the weak predictors. Moving averages and granularities are expressed in hours (h), the number of GCR observations preceding the studied one are indicated by absolute numbers (n) and the time window length is expressed in days (d).}\label{tab:datasets}
\end{table}

In order to develop and test the reliability of an ML model providing solar wind speed predictions, GCR data gathered by the LPF mission and solar wind speed measurements of the ACE spacecraft instruments in L1 in 2016--2017 were collected and binned hourly to keep the statistical fluctuations below 1\%.
The observation time series were aligned in time to build the input data set.
%
%
It is worthwhile to notice that the GCR data are expressed in percent variations of the GCR flux w.r.t.\ the average value during each Bartels rotation (BR) from mid-February 2016 through the beginning of July 2017.
We recall here that BRs are periods of 27 days representing complete apparent Sun rotations viewed from Earth.
The BRs are enumerated since February 8, 1832, being this the day 1 of the BR 1.
BRs 2491--2508 are studied in this work.
In other words, LPF provided GCR data during 18 complete BRs.
Since the LPF GCR time series presented some missing data, the neural network model described in~\cite{Villani22} was used to fill these gaps.

To obtain a model with good predictive capabilities it is necessary to optimise \textit{(i)} the data pre-processing routine, \textit{(ii)} the ML model, and \textit{(iii)} the hyper-parameter values.
It is widely recognised that ML ensemble models present a better accuracy than single, weaker predictors~\citep{ens1,ens2}.
For this reason, the presented work is based on an ensemble regressor composed of 28 base models belonging to 5 different categories: linear regressors (LRs), $k$-nearest neighbours ($k$-NN) models, support vector machines (SVMs), random forests (RFs) and artificial neural networks (ANNs).
All categories fall inside the supervised learning paradigm.
These models have been chosen because they provided good predictions in some subsets of the data and an ensemble model combining all models returned better outputs in the whole test set.

\subsection{Data pre-processing and data set selection}\label{ssec:prep}

The GCR intensity in the interplanetary medium is modulated by the passage of interplanetary magnetic structures and solar wind HSS.
In particular, recurrent GCR flux variations are associated with the transit of solar wind HSS originated from coronal holes.
The recurrent GCR flux depressions present periodicities correlated with the Sun rotation period and higher harmonics.

For the model presented here an appropriate data set of the GCR flux variations measured with LPF was built accordingly.
Each instance of the data set is composed of a solar wind output data point associated with a set of contemporaneous and preceding GCR flux observations.
The number of selected preceding GCR data points ($n$, henceforth) is a critical factor for the final predictive performance of the model and it was object of detailed investigations.
In order to limit the amount of input features of the model, only one GCR data point every $g$ hours in the LPF GCR time series was used.
In the following, this parameter is referred to as the observation \textit{granularity}.
Also the observation granularity is an important parameter to be analysed and tuned, strongly impacting the model performance.
Granularities from 1 to 6 hours were therefore tested during a preliminary study.
Each different granularity value was tested in combination with 5 values for the number of preceding GCR data points (in particular, $n \in \{ 24, 36, 48, 72, 96 \}$), resulting in 30 overall distinct data sets to be evaluated.
GCR flux and solar wind speed time series of each data set were smoothed via a moving average of $g$ hours, equal to the granularity value.

Several tests were carried out before the ML model training to select the best values for $n$ and $g$.
Results showed data sets having $n \in \{24, 48\}$ and $g \in \{3, 5\}$ as the most promising.
Four data sets were selected accordingly with the optimised parameters summarised in \Cref{tab:datasets}.
Here and in the following, the four data sets are indicated by DS$i$, with $i \in \{1, ..., 4\}$.
\Cref{tab:datasets} also reports the time window length for the LPF GCR data, obtained by multiplying $n$ observations taken every $g$ hours.

\subsection{Weak predictors}\label{ssec:pred}

Our final regressor is composed of 7 different models, each one cloned 4 times and trained with each of the four data sets reported in \Cref{tab:datasets}.
The ensemble predictor is thus based on $7 \times 4 = 28$ weak regressors and provides its outputs by averaging the whole set of predictions.

All models are trained and queried with scaled data sets.
The scaling phase takes into considerations both mean and standard deviation of the input and output features and its goal is to strengthen the model training.

The data set split into training, validation and test set follows the same rationale for every model.
Since the overall data set contains observations gathered during 18 complete BRs, 1 of them was selected as test set, 2 of them as validation set and the remaining 15 as training set, resulting in 648 test instances, 1\,296 validation instances and 9\,720 training instances. 
Only the training set was adopted to train the model.
The validation set had the only purpose to select the best model hyper-parameter values, whereas the test set was used to assess the model predictive performance on completely new and unknown instances w.r.t.\ the training and validation sets.

We repeated for 18 times the model training with the described data splitting strategy in order to use every end each BR as test set.
Each weak model was consequently tested 18 times.
So it was the ensemble model.
We considered this approach since each BR has its own peculiarities and it is important to spot in which cases a predictor has a poor performance, in order to possibly solve the problem.

\subsubsection{Linear regressors}

The simplest weak models composing the presented ensemble predictor are LRs, expressing the data set output feature as a linear combination of the input variables.
The training phase of these models consists in the minimisation of the residual sum of squares calculated between the data set output feature and the predictions provided by the linear models.
No hyper-parameters are required for this kind of predictor.

In \Cref{tab:lr} the LR predictive performance expressed as mean absolute error (MAE) between the solar wind speed model predictions and the corresponding expected outputs is reported for all BRs and for each data set described in \Cref{tab:datasets}.
MAE values are expressed in km~s$^{-1}$.
MAE corresponding to the predictions obtained by averaging the LRs applied to the 4 data sets is added in the rightmost column.
%
%
The MAE averaged over all BRs is reported in the bottom row.

\begin{figure}
	\centering
	\includegraphics[width=\wid]{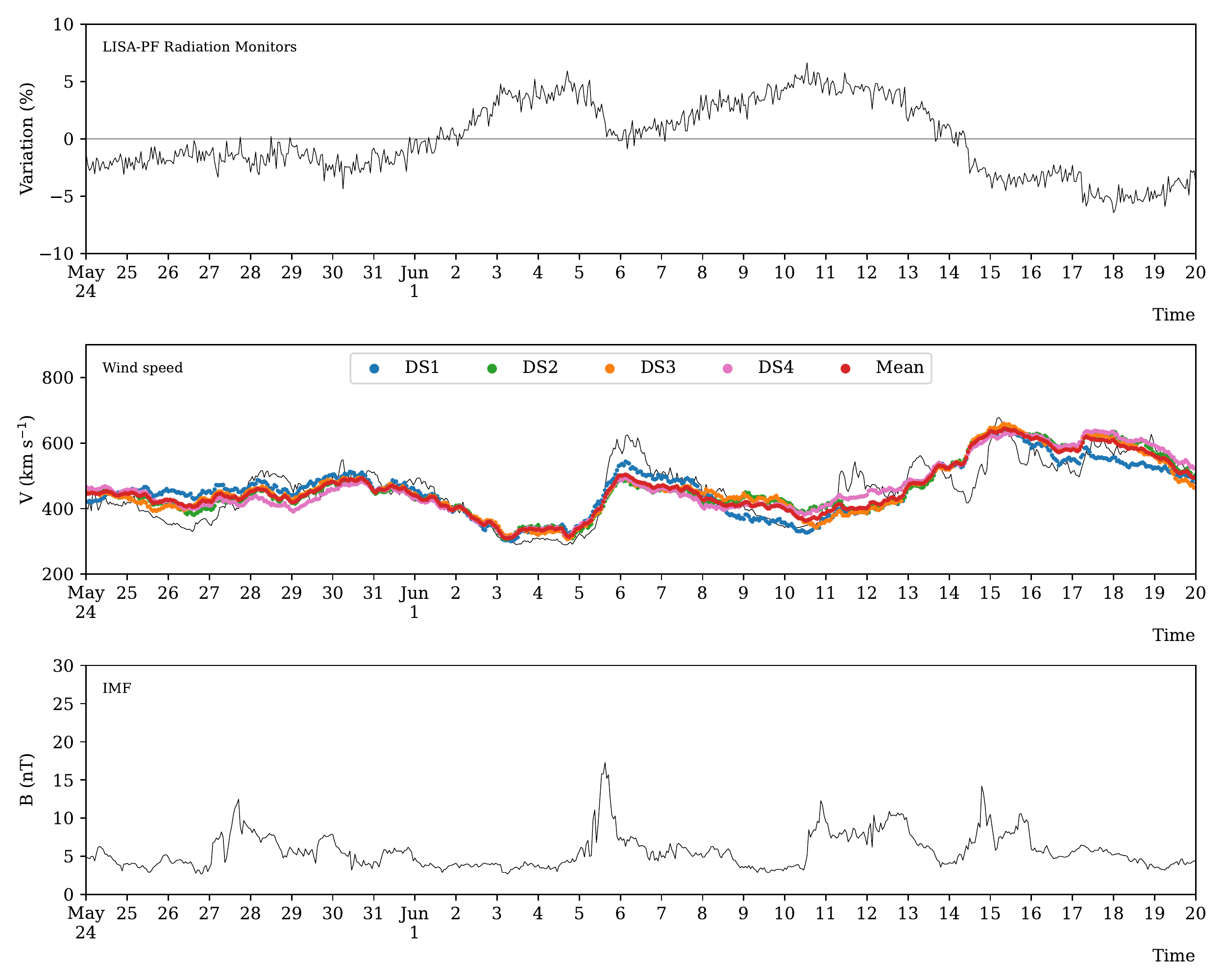}
	\caption{Inputs and outputs of the LR models for the BR 2494 (from May 24 through June 20, 2016). The LPF GCR observations used as input data for the LR models are reported in the top panel. Solar wind speed data gathered by ACE (black line) and predictions provided by LR weak models for each one of the data sets reported in \Cref{tab:datasets} (coloured dots) are shown in the middle panel. In the same panel the average predictions obtained by considering the 4 data sets are depicted in red. The IMF intensity is also shown in the bottom panel.}\label{fig:lr}
\end{figure}

A graphical example of the solar wind speed reconstruction provided only by the LRs for the BR 2495 (from May 24 through June 20, 2016) is reported in \Cref{fig:lr}.
We point out that predictive performance quantitative assessments and comparisons between the adopted weak models can be carried out on the basis of the data reported in the tables. Figures have the only purpose of showing the agreement between solar wind speed predictions and data from a qualitative perspective. Thus, we show for each weak model the corresponding predictions given for a different BR in order to demonstrate the effectiveness of our models in several partitions of the data set.

In the top panel of \Cref{fig:lr} the LPF GCR flux percent variations are reported for the BR 2495.
Contemporaneous observations of the solar wind speed and of the IMF intensity gathered by the ACE experiment are reported in the middle and bottom panels, respectively.
LR model predictions for the 4 data sets described in \Cref{tab:datasets} are also shown in the middle panel as coloured dots.
Finally, the average predictions obtained with LR models by considering the 4 data sets are reported as red dots.
It is worthwhile to notice that IMF parameters are not taken into account to carry out solar wind speed predictions.
Nevertheless, the IMF intensity increase above 10~nT reveals the passage of interplanetary magnetic structures, at the origin of GCR flux \textit{non-recurrent} variations, and of corotating interaction regions observed at the interface between slow and fast wind.
Since the presented models are trained to provide a solid solar wind speed reconstruction during GCR recurrent variations at the time LPF remained into orbit, it is important to include in the plot also the magnetic field intensity to possibly identify the passage of ICMEs w.r.t.\ the transit of solar wind HSS.
More in general, during periods of IMF intensities larger than 10~nT the GCR particle drift in the magnetic field overcomes the effect of convection in the solar wind and the effect of the IMF increase must be taken into account to correctly evaluate the model performance.

\begin{table}
	\begin{tabular}{cccccc@{\hskip 0.5in}cccccc}
		\toprule
		BR & DS1 & DS2 & DS3 & DS4 & Mean & BR & DS1 & DS2 & DS3 & DS4 & Mean \\
		\midrule
		2491 & 59.05 & 56.25 & 58.12 & 56.26 & 56.43 & 2500 & 83.77 & 71.48 & 73.38 & 70.77 & 74.37 \\
		2492 & 63.81 & 57.89 & 59.63 & 58.93 & 58.29 & 2501 & 94.82 & 81.66 & 83.93 & 79.53 & 84.85 \\
		2493 & 61.09 & 61.84 & 59.77 & 60.77 & 59.64 & 2502 & 93.67 & 92.84 & 91.03 & 94.33 & 91.95 \\
		2494 & 38.24 & 42.27 & 40.43 & 41.97 & 38.37 & 2503 & 81.12 & 85.54 & 85.93 & 85.84 & 84.06 \\
		2495 & 44.39 & 51.11 & 48.09 & 48.87 & 47.04 & 2504 & 83.24 & 88.45 & 87.89 & 86.94 & 86.34 \\
		2496 & 78.52 & 63.47 & 63.91 & 64.78 & 66.39 & 2505 & 78.15 & 81.34 & 78.81 & 81.21 & 79.15 \\
		2497 & 68.47 & 57.55 & 57.99 & 59.00 & 58.99 & 2506 & 82.62 & 79.68 & 80.09 & 79.21 & 79.36 \\
		2498 & 86.44 & 75.79 & 78.41 & 70.45 & 77.46 & 2507 & 73.24 & 70.03 & 70.14 & 65.78 & 69.11 \\
		2499 & 75.08 & 67.83 & 68.79 & 64.95 & 68.29 & 2508 & 67.86 & 70.11 & 69.79 & 70.54 & 69.36 \\
		\midrule
		& & & & & & Mean & 72.98 & 69.73 & 69.78 & 68.90 & 69.41 \\
		\bottomrule
	\end{tabular}
	\caption{MAE for each BR and for each data set appearing in \Cref{tab:datasets} calculated for the LR. The mean MAE over the whole set of BRs is reported in the last row. MAE values are expressed in km~s$^{-1}$.}\label{tab:lr}
\end{table}

\subsubsection{$k$-nearest neighbours}

$k$-NN regressors are simple models that, when queried with an instance, provide a prediction by averaging the output value of the $k$ nearest training instances (i.e., the \textit{neighbours}) w.r.t.\ the query.
The contribution of each neighbour may be uniform or weighted.
In our experiments we tested several approaches with different $k$ values.
More in detail, $k$ values ranging from 1 to 250 were examinated.
Each $k$-NN was tested with both uniform and weighted neighbour contributions.
We adopted weighted neighbour contributions according to the distance between neighbours and query.
Distance between instances may be calculated according to various strategies.
We chose to limit our experiments to the Manhattan and Euclidean distances.

\begin{figure}
	\centering
	\includegraphics[width=\wid]{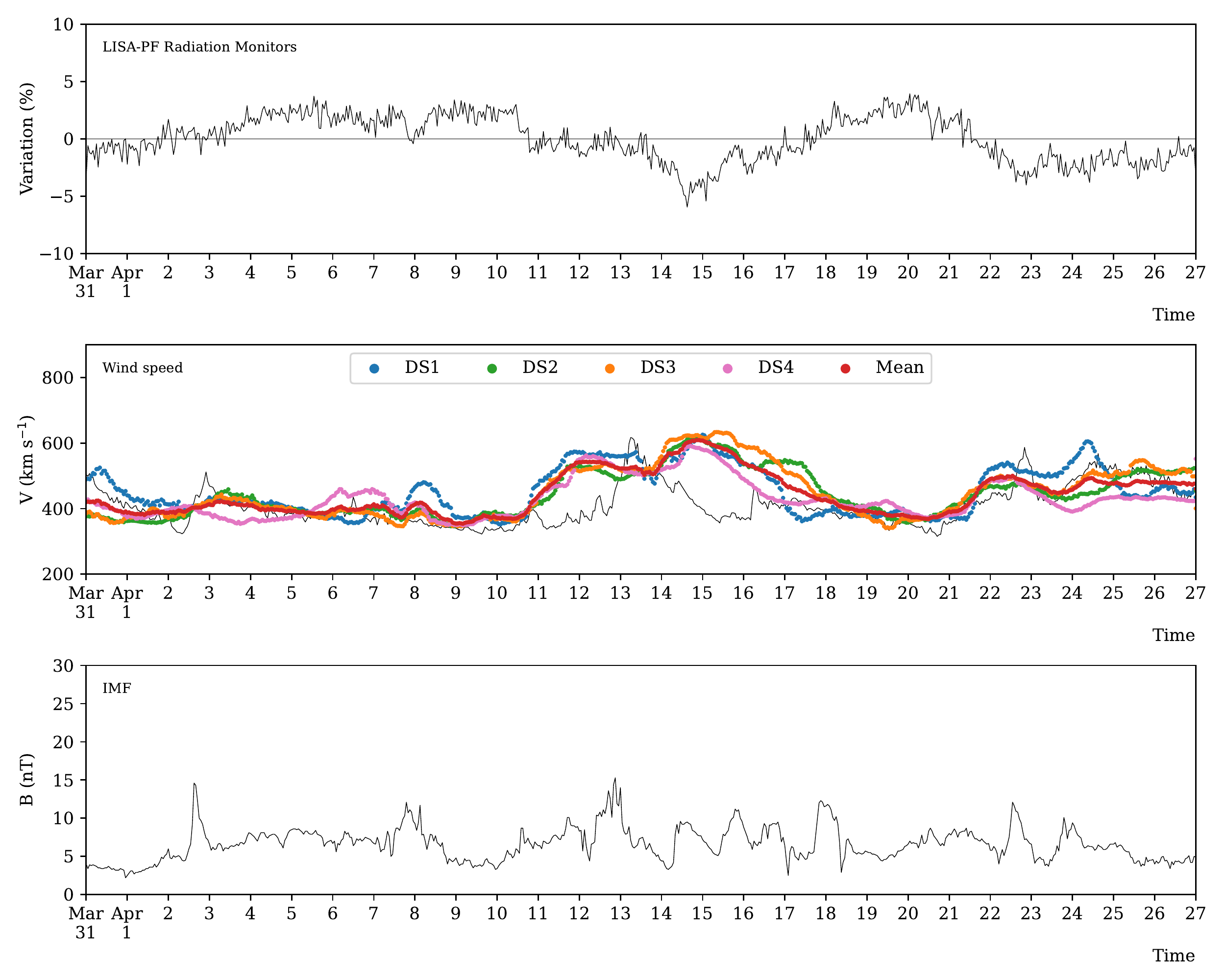}
	\caption{Same as \Cref{fig:lr} for the 100-NNs reproducing the solar wind speed during the BR 2492 (from March 31 through April 27, 2016).}\label{fig:knn}
\end{figure}

From our experiments the following hyper-parameters resulted those correlated with the best model outputs: $k = 100$, uniform neighbour contribution, and Euclidean distance.
The MAE associated to this optimum model predictions of the solar wind speed are reported in \Cref{tab:knn}, following the same schema adopted for \Cref{tab:lr}.
Predictions provided by the optimum models are reported in \Cref{fig:knn} for the BR 2492 (from March 31 through April 27, 2016), as an example.
Figure panels have the same meaning as in \Cref{fig:lr}.

\begin{table}
	\begin{tabular}{cccccc@{\hskip 0.5in}cccccc}
		\toprule
		BR & DS1 & DS2 & DS3 & DS4 & Mean & BR & DS1 & DS2 & DS3 & DS4 & Mean \\
		\midrule
		2491 & 64.42 & 55.63 & 59.62 & 57.08 & 55.79 & 2500 & 81.04 & 58.87 & 58.93 & 57.60 & 60.91 \\
		2492 & 58.41 & 55.25 & 54.32 & 54.80 & 49.73 & 2501 & 93.51 & 81.75 & 78.59 & 70.41 & 80.16 \\
		2493 & 66.76 & 64.92 & 67.23 & 72.29 & 61.09 & 2502 & 97.51 & 94.96 & 92.91 & 112.97 & 97.46 \\
		2494 & 56.56 & 53.79 & 48.78 & 56.32 & 45.52 & 2503 & 71.21 & 69.81 & 67.89 & 61.70 & 61.90 \\
		2495 & 57.75 & 55.52 & 48.79 & 58.20 & 49.73 & 2504 & 77.79 & 73.16 & 72.61 & 56.84 & 67.17 \\
		2496 & 72.65 & 52.83 & 56.41 & 55.24 & 54.63 & 2505 & 81.53 & 88.73 & 89.91 & 106.34 & 89.37 \\
		2497 & 76.48 & 68.40 & 65.52 & 69.19 & 66.11 & 2506 & 96.73 & 88.29 & 88.20 & 83.33 & 86.73 \\
		2498 & 83.24 & 77.77 & 79.90 & 80.83 & 76.65 & 2507 & 69.94 & 70.74 & 67.78 & 69.53 & 65.13 \\
		2499 & 78.62 & 70.26 & 75.46 & 73.19 & 69.87 & 2508 & 72.55 & 79.17 & 74.45 & 66.07 & 68.81 \\
		\midrule
		& & & & & & Mean & 75.37 & 69.99 & 69.30 & 70.11 & 67.04 \\
		\bottomrule
	\end{tabular}
	\caption{Same as \Cref{tab:lr} for the 100-NN models.}\label{tab:knn}
\end{table}

\subsubsection{Support vector machines}

\begin{table}
	\begin{tabular}{cccccc@{\hskip 0.5in}cccccc}
		\toprule
		BR & DS1 & DS2 & DS3 & DS4 & Mean & BR & DS1 & DS2 & DS3 & DS4 & Mean \\
		\midrule
		2491 & 58.77 & 56.76 & 59.17 & 69.34 & 57.14 & 2500 & 81.38 & 63.47 & 68.83 & 61.06 & 67.85 \\
		2492 & 57.93 & 51.71 & 53.16 & 64.25 & 51.06 & 2501 & 87.53 & 74.43 & 76.49 & 73.33 & 77.34 \\
		2493 & 69.52 & 74.39 & 73.62 & 80.26 & 70.41 & 2502 & 100.09 & 94.23 & 89.55 & 100.70 & 94.93 \\
		2494 & 57.53 & 47.26 & 47.94 & 54.75 & 43.06 & 2503 & 71.03 & 77.98 & 77.46 & 73.90 & 74.86 \\
		2495 & 46.18 & 46.33 & 44.53 & 54.66 & 43.40 & 2504 & 79.31 & 83.68 & 81.62 & 71.54 & 78.31 \\
		2496 & 68.89 & 45.42 & 53.60 & 50.95 & 52.36 & 2505 & 86.76 & 91.30 & 85.27 & 93.50 & 87.69 \\
		2497 & 75.23 & 58.10 & 62.77 & 61.45 & 60.61 & 2506 & 91.58 & 90.04 & 88.70 & 84.03 & 86.43 \\
		2498 & 85.33 & 78.71 & 80.22 & 70.73 & 77.15 & 2507 & 69.55 & 67.80 & 66.71 & 78.98 & 68.51 \\
		2499 & 76.45 & 76.47 & 73.33 & 81.20 & 75.21 & 2508 & 65.54 & 61.65 & 63.80 & 64.27 & 63.07 \\
		\midrule
		& & & & & & Mean & 73.81 & 68.87 & 69.26 & 71.61 & 68.30 \\
		\bottomrule
	\end{tabular}
	\caption{Same as \Cref{tab:lr} for the SVM models.}\label{tab:svm}
\end{table}

\begin{figure}
	\centering
	\includegraphics[width=\wid]{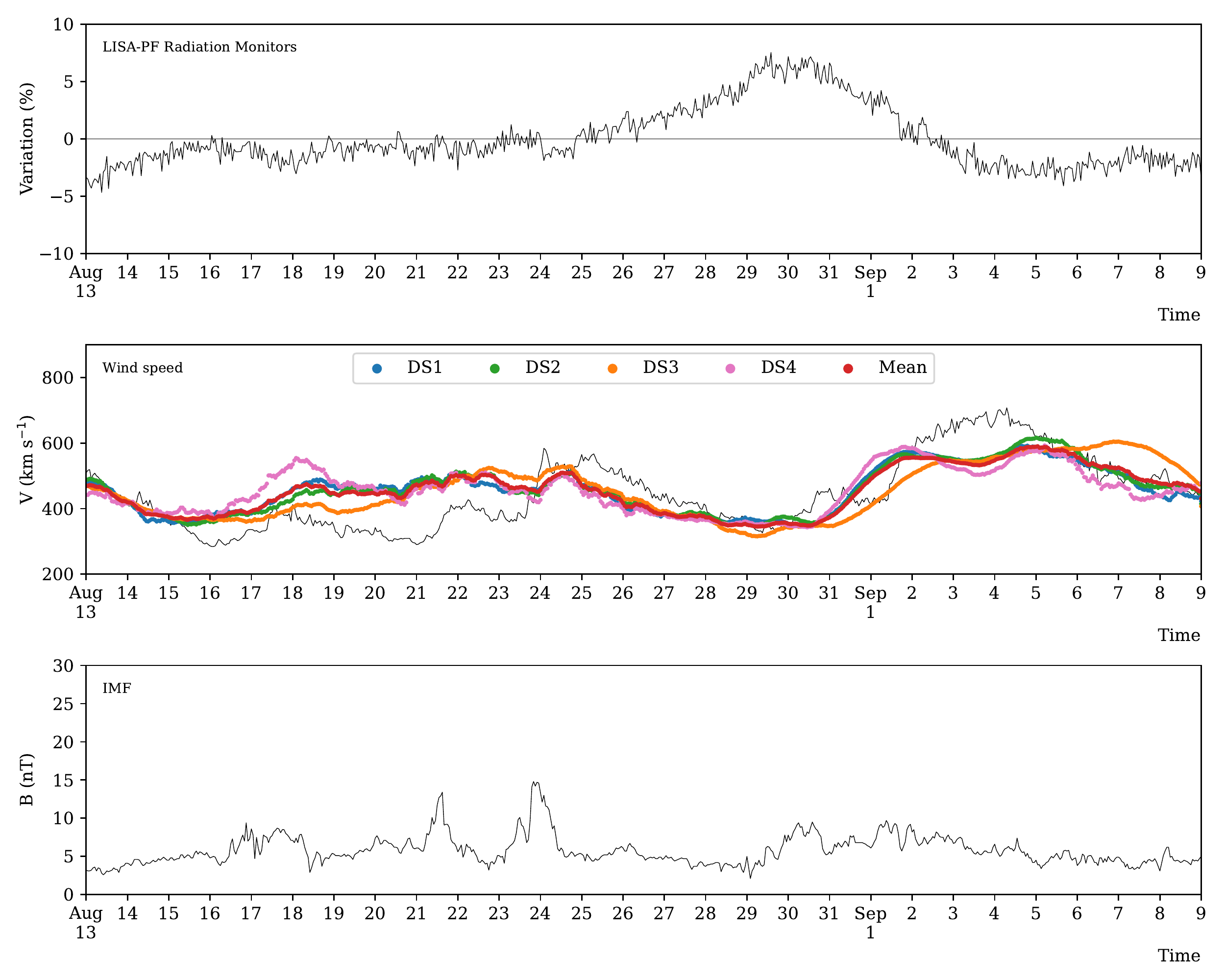}
	\caption{Same as \Cref{fig:lr} for the SVMs reproducing the solar wind speed during the BR 2497 (from August 13 through September 9, 2016).}\label{fig:svm}
\end{figure}

SVM models applied to regression tasks provide predictions after having fitted a hyper-plane during the training phase, similarly to the classification counterparts~\citep{svm2,svm1}.
Tuning the hyper-parameters of a SVM usually means choosing the best kernel function and the optimum values for the regularisation parameter and the maximum error $\varepsilon$.

As for the kernel function, SVM adopting radial basis function (RBF) kernels proved to have better performances than other SVM with different kernels, as for instance linear or polynomial.
We recall here that RBF kernels are amongst the most used ones, allowing to map a generic data set, possibly nonlinearly separable, into another data set with higher dimensionality but linearly separable.
Meticulous experiments about the best values for $\varepsilon$ and the regularisation parameter were carried out.
In particular, $\varepsilon$ controls the model sensitivity, since during the training phase predictions correlated to errors smaller than $\varepsilon$ are not penalised.
A very small value for $\epsilon$ hinders the generalisation capability of the SVM.
We found $\varepsilon=0.05$ a suitable choice.
Finally, for the regularisation parameter several values ranging between 0.01 and 5 were tested, highlighting 0.1 as an optimum value for the task at hand.

The predictive error of the SVMs trained with the described hyper-parameters and applied to the data sets reported in \Cref{tab:datasets} is shown in \Cref{tab:svm}.
\Cref{fig:svm} depicts the solar wind speed reconstruction obtained with the SVM models for the BR 2497 (from August 13 through September 9, 2016).

\begin{table}
	\begin{tabular}{cccccc@{\hskip 0.5in}cccccc}
		\toprule
		BR & DS1 & DS2 & DS3 & DS4 & Mean & BR & DS1 & DS2 & DS3 & DS4 & Mean \\
		\midrule
		2491 & 65.51 & 61.68 & 65.86 & 63.51 & 59.62 & 2500 & 85.88 & 66.09 & 73.26 & 66.99 & 71.89 \\
		2492 & 62.53 & 68.07 & 67.72 & 67.14 & 58.69 & 2501 & 96.39 & 83.44 & 88.29 & 75.59 & 85.17 \\
		2493 & 72.86 & 72.14 & 78.29 & 78.82 & 69.08 & 2502 & 93.50 & 94.22 & 93.72 & 94.67 & 91.55 \\
		2494 & 67.54 & 62.54 & 62.88 & 49.61 & 51.27 & 2503 & 71.71 & 66.47 & 73.16 & 58.85 & 63.88 \\
		2495 & 56.24 & 53.59 & 50.59 & 47.66 & 44.05 & 2504 & 79.07 & 81.13 & 78.15 & 62.96 & 72.59 \\
		2496 & 70.54 & 51.02 & 58.47 & 56.68 & 53.46 & 2505 & 81.55 & 84.64 & 85.38 & 89.55 & 82.53 \\
		2497 & 80.65 & 72.28 & 75.62 & 71.17 & 70.67 & 2506 & 88.15 & 89.88 & 91.77 & 87.28 & 86.47 \\
		2498 & 86.44 & 74.37 & 79.91 & 75.01 & 76.03 & 2507 & 62.55 & 64.42 & 56.82 & 72.07 & 59.33 \\
		2499 & 81.22 & 72.81 & 73.60 & 77.16 & 71.85 & 2508 & 72.33 & 70.80 & 69.72 & 66.51 & 66.90 \\
		\midrule
		& & & & & & Mean & 76.37 & 71.64 & 73.51 & 70.07 & 68.61 \\
		\bottomrule
	\end{tabular}
	\caption{Same as \Cref{tab:lr} for the RF models.}\label{tab:rf}
\end{table}

\begin{figure}
	\centering
	\includegraphics[width=\wid]{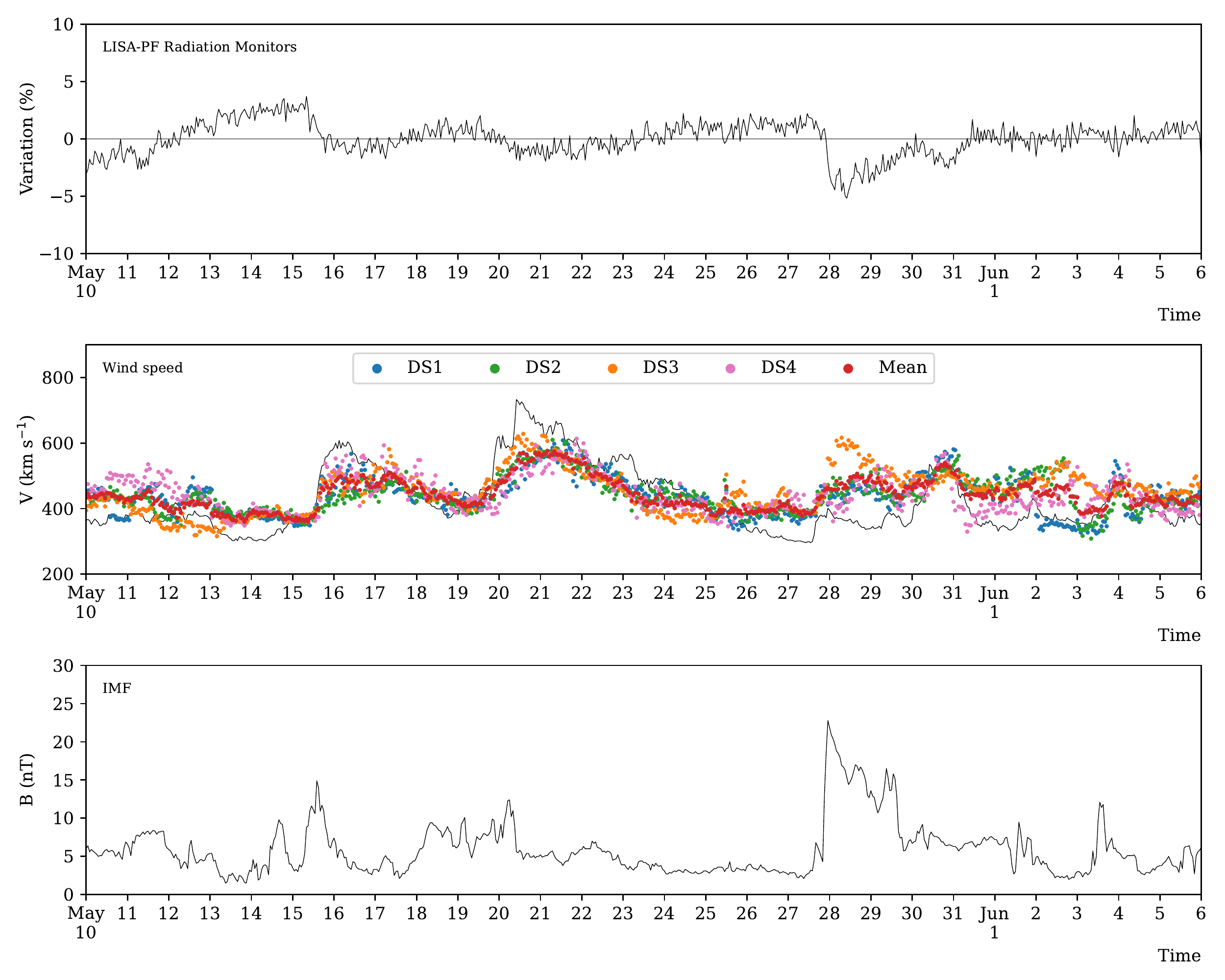}
	\caption{Same as \Cref{fig:lr} for the RFs reproducing the solar wind speed during the BR 2507 (from May 10 through June 6, 2017).}\label{fig:rf}
\end{figure}

\subsubsection{Random forests}

RFs are ensemble predictors based on underlying decision trees, each one provided with the whole training set or a subset of it~\citep{breiman2001random}.
As other ensemble models, their outputs are the average value of those given by the base predictors and, consequently, they show a better accuracy than others.

The hyper-parameters of these models are the number of base decision trees to be trained and, in addition, those required by the base decision trees.
In our experiments 5 to 100 base trees were tested.
Unbounded trees were adopted, i.e., decision trees without constraints on the maximum depth or leaf amount.
The best solar wind speed reconstruction was obtained with RFs having 20 base trees.
The corresponding prediction MAE is reported in \Cref{tab:rf}.
As an example, the solar wind speed reconstruction for the BR 2507 (from May 10 through June 6, 2017) is shown in \Cref{fig:rf}.

\subsubsection{Artificial neural networks}

Neural networks are amongst the most promising models adopted as predictors for both classification and regression, since they can learn arbitrarily complex input/output relationships~\citep{ann1,ann2}.
However, this kind of model requires the fine tuning of a large set of hyper-parameters and also the training phase is subject to a number of critical choices.
For this particular regression task, we opted for a model training with early stopping triggered by the predictive error measured on the validation set.
Several patience values were tested for the early stopping monitoring, selecting 20 as the best one.
The maximum number of training iterations was fixed to 300, allowing slow trainings to reach convergence.

\begin{table}[]
	\begin{tabular}{cccc}
		\toprule
		Activation function & Hidden neurons & Learning rate & Batch size \\
		\midrule
		Sigmoid & 500; 200 & 0.005 & 1000 \\
		Rectified linear & 500; 200 & 0.01 & 500 \\
		Hyperbolic tangent & 500; 200 & 0.01 & 1000 \\
		\bottomrule
	\end{tabular}
	\caption{Hyper-parameters of the 3 ANN models having the best predictive performance and selected in the weak regressor pool.}\label{tab:hyp}
\end{table}

\begin{figure}[]
	\centering
	\includegraphics[width=\wid]{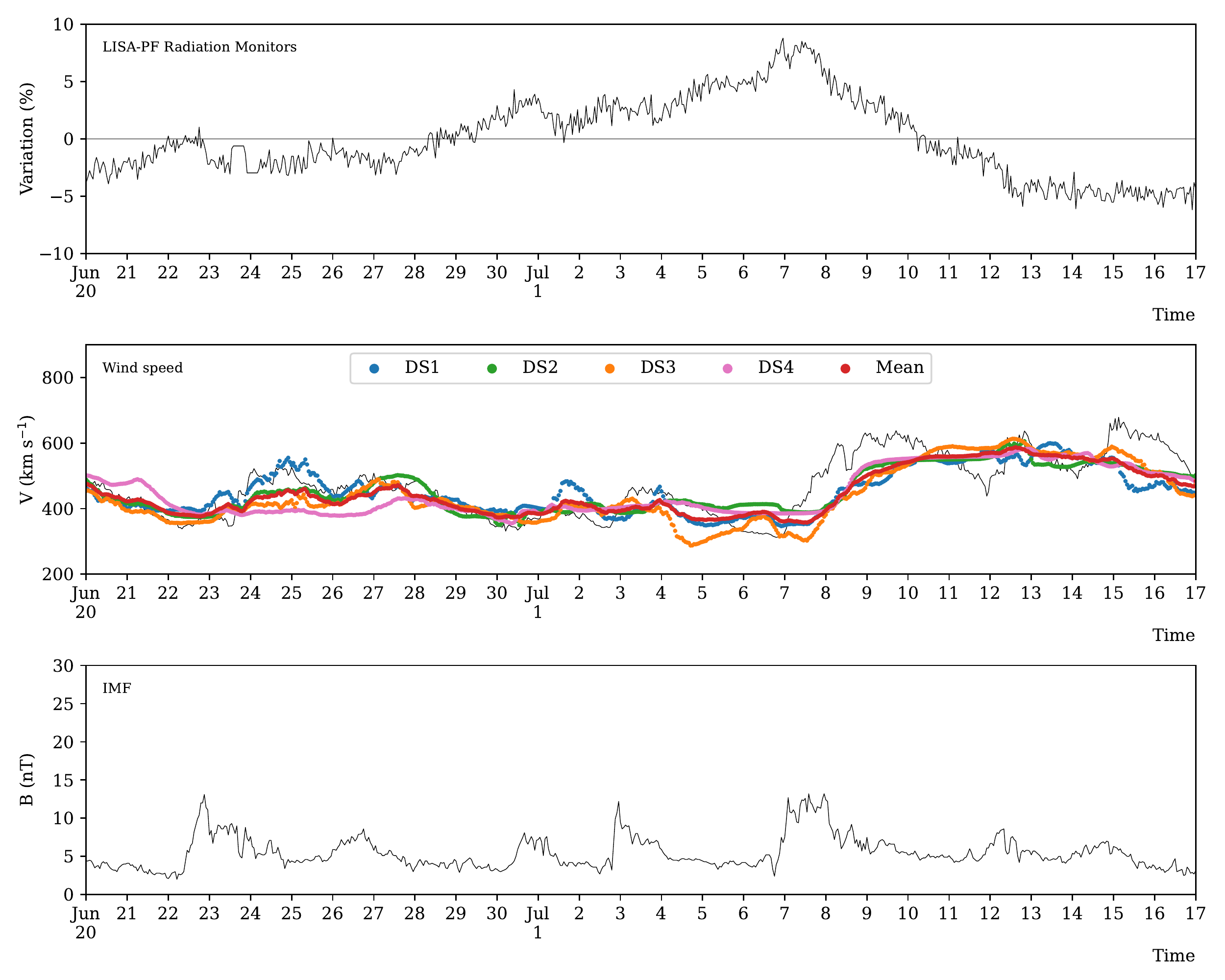}
	\caption{Same as \Cref{fig:lr} for the neural network having the hyperbolic tangent activation function for the hidden neurons. The solar wind speed trend is reproduced during the BR 2495 (from June, 20 through July 17, 2016).}\label{fig:ann}
\end{figure}

\begin{table}[]
	\begin{tabular}{cccccc@{\hskip 0.5in}cccccc}
		\toprule
		BR & DS1 & DS2 & DS3 & DS4 & Mean & BR & DS1 & DS2 & DS3 & DS4 & Mean \\
		\midrule
		2491 & 66.31 & 62.62 & 85.83 & 52.70 & 62.33 & 2500 & 73.79 & 75.91 & 70.10 & 60.07 & 68.63 \\
		2492 & 63.78 & 64.57 & 64.70 & 60.55 & 58.30 & 2501 & 92.32 & 77.64 & 70.81 & 83.90 & 79.09 \\
		2493 & 64.31 & 67.02 & 69.88 & 67.59 & 62.60 & 2502 & 99.59 & 90.40 & 87.47 & 100.42 & 91.96 \\
		2494 & 45.46 & 52.31 & 44.20 & 54.04 & 42.59 & 2503 & 67.74 & 81.34 & 77.29 & 77.24 & 74.49 \\
		2495 & 49.54 & 40.75 & 50.66 & 50.15 & 43.82 & 2504 & 72.31 & 80.93 & 82.84 & 83.00 & 75.97 \\
		2496 & 73.86 & 52.33 & 54.95 & 61.74 & 55.84 & 2505 & 74.89 & 84.37 & 75.42 & 84.88 & 78.72 \\
		2497 & 69.59 & 67.56 & 55.63 & 58.69 & 59.33 & 2506 & 89.73 & 90.35 & 98.01 & 85.10 & 87.95 \\
		2498 & 89.93 & 72.30 & 76.64 & 72.57 & 76.57 & 2507 & 72.10 & 67.01 & 60.77 & 84.83 & 66.77 \\
		2499 & 77.19 & 64.06 & 72.75 & 65.36 & 67.43 & 2508 & 82.05 & 64.02 & 66.13 & 64.41 & 66.62 \\
		\midrule
		& & & & & & Mean & 73.58 & 69.75 & 70.23 & 70.40 & 67.72 \\
		\bottomrule
	\end{tabular}
	\caption{Same as \Cref{tab:lr} for the ANN models with hyperbolic tangent activation function.}\label{tab:tanh}
\end{table}

\begin{table}
	\begin{tabular}{cccccc@{\hskip 0.5in}cccccc}
		\toprule
		BR & DS1 & DS2 & DS3 & DS4 & Mean & BR & DS1 & DS2 & DS3 & DS4 & Mean \\
		\midrule
		2491 & 64.71 & 56.54 & 65.44 & 73.54 & 59.14 & 2500 & 82.40 & 67.97 & 68.90 & 54.85 & 65.66 \\
		2492 & 70.84 & 58.54 & 69.68 & 70.71 & 61.74 & 2501 & 88.38 & 84.85 & 84.44 & 75.31 & 82.31 \\
		2493 & 64.43 & 73.78 & 77.80 & 86.29 & 70.05 & 2502 & 93.33 & 82.30 & 83.39 & 104.56 & 88.89 \\
		2494 & 43.62 & 48.52 & 40.52 & 58.89 & 39.20 & 2503 & 73.63 & 68.16 & 66.73 & 70.59 & 65.97 \\
		2495 & 53.09 & 54.86 & 47.79 & 59.33 & 49.10 & 2504 & 84.47 & 85.61 & 78.84 & 71.17 & 78.58 \\
		2496 & 71.99 & 69.41 & 71.39 & 72.55 & 67.30 & 2505 & 70.79 & 81.56 & 82.61 & 95.69 & 81.37 \\
		2497 & 70.98 & 50.78 & 57.78 & 50.99 & 53.56 & 2506 & 96.82 & 94.42 & 103.17 & 85.27 & 90.19 \\
		2498 & 88.39 & 74.55 & 75.73 & 60.24 & 71.58 & 2507 & 76.87 & 75.96 & 74.27 & 66.46 & 72.57 \\
		2499 & 77.12 & 62.89 & 61.35 & 88.09 & 68.15 & 2508 & 67.78 & 68.05 & 68.88 & 66.45 & 67.22 \\
		\midrule
		& & & & & & Mean & 74.42 & 69.93 & 71.04 & 72.83 & 68.48 \\
		\bottomrule
	\end{tabular}
	\caption{Same as \Cref{tab:lr} for the ANN models with rectified linear activation function.}\label{tab:relu}
\end{table}

\begin{table}
	\begin{tabular}{cccccc@{\hskip 0.5in}cccccc}
		\toprule
		BR & DS1 & DS2 & DS3 & DS4 & Mean & BR & DS1 & DS2 & DS3 & DS4 & Mean \\
		\midrule
		2491 & 61.31 & 57.06 & 62.28 & 59.25 & 52.35 & 2500 & 83.12 & 70.10 & 71.38 & 61.63 & 69.04 \\
		2492 & 60.21 & 47.49 & 47.70 & 53.31 & 47.88 & 2501 & 91.79 & 78.83 & 70.27 & 76.10 & 78.99 \\
		2493 & 62.04 & 60.66 & 64.38 & 61.47 & 56.17 & 2502 & 103.08 & 87.95 & 85.93 & 106.09 & 94.35 \\
		2494 & 42.42 & 57.01 & 39.21 & 62.81 & 44.38 & 2503 & 75.10 & 94.88 & 75.33 & 97.98 & 84.80 \\
		2495 & 50.81 & 53.66 & 48.61 & 46.59 & 46.08 & 2504 & 77.90 & 82.08 & 75.71 & 84.17 & 78.63 \\
		2496 & 71.33 & 84.56 & 63.00 & 69.96 & 65.34 & 2505 & 79.67 & 91.44 & 92.41 & 106.40 & 88.87 \\
		2497 & 74.11 & 64.18 & 61.17 & 60.86 & 62.71 & 2506 & 81.99 & 89.14 & 96.62 & 71.34 & 82.77 \\
		2498 & 86.05 & 85.65 & 74.25 & 88.16 & 82.43 & 2507 & 69.22 & 65.39 & 64.12 & 72.77 & 65.89 \\
		2499 & 74.17 & 66.31 & 66.67 & 74.38 & 68.32 & 2508 & 65.72 & 65.98 & 65.07 & 71.09 & 66.67 \\
		\midrule
		& & & & & & Mean & 72.78 & 72.35 & 68.01 & 73.58 & 68.65 \\
		\bottomrule
	\end{tabular}
	\caption{Same as \Cref{tab:lr} for the ANN models with sigmoid activation function.}\label{tab:sigm}
\end{table}

Other important hyper-parameters carefully investigated for this work were the batch size, the learning rate, the network architecture and the neurons' activation function.
Since these parameters are usually intertwined, a grid search was performed to highlight the most promising combination of values to be assigned to the hyper-parameters, instead of exploring the admissible value space for each individual parameter.
Batch sizes from 100 to 1000 were tested, whereas for the learning rate values ranging between 0.001 and 0.05 were adopted.
As for the ANN architecture, a multilayer perceptron (MLP) was chosen, i.e., a fully connected neural network relying on hidden layers of artificial neurons.
Several architectures having 2, 3 or 4 hidden layers were investigated, with different amounts of hidden neurons in each layer.
The overall number of neurons varied from 150 to 1800 and in all the experiments the amount of neurons in the deepest layers was always kept smaller than those in the shallowest ones.
Finally, 3 different activation functions were tested for hidden neurons: hyperbolic tangent, sigmoid and rectified linear.

The presented ensemble regressor relies on 3 classes of ANNs, since the best MLP with each one of the aforementioned activation functions was chosen as concurring base model.
The 4-layered architecture is characterised by 2 inner hidden layers.
The best average predictive performance was achieved by selecting 500 neurons for the first hidden layer and 200 neurons for the second.
Learning rate and batch size best values are different for each base model.
In particular, the neural network with sigmoid activation function showed the best results with a learning rate equal to 0.005, whereas for the other 2 a value equal to 0.01 was found more suitable.
A batch size of 500 was adopted for the MLP with rectified linear activations, while 100 was the choice for the remaining 2.
The best hyper-parameter values for the ANNs selected as weak regressors are shown in \Cref{tab:hyp}.
The prediction MAE of the 3 best models is reported in \Cref{tab:tanh,tab:relu,tab:sigm}.
A graphical predictive example is depicted only for the ANN with hyperbolic tangent activation function in \Cref{fig:ann} for solar wind speed reconstruction during the BR 2495 (from June, 20 through July 17, 2016).

\subsubsection{Weak model outcomes}

\begin{figure}[t]
	\centering
	\includegraphics[width=\linewidth]{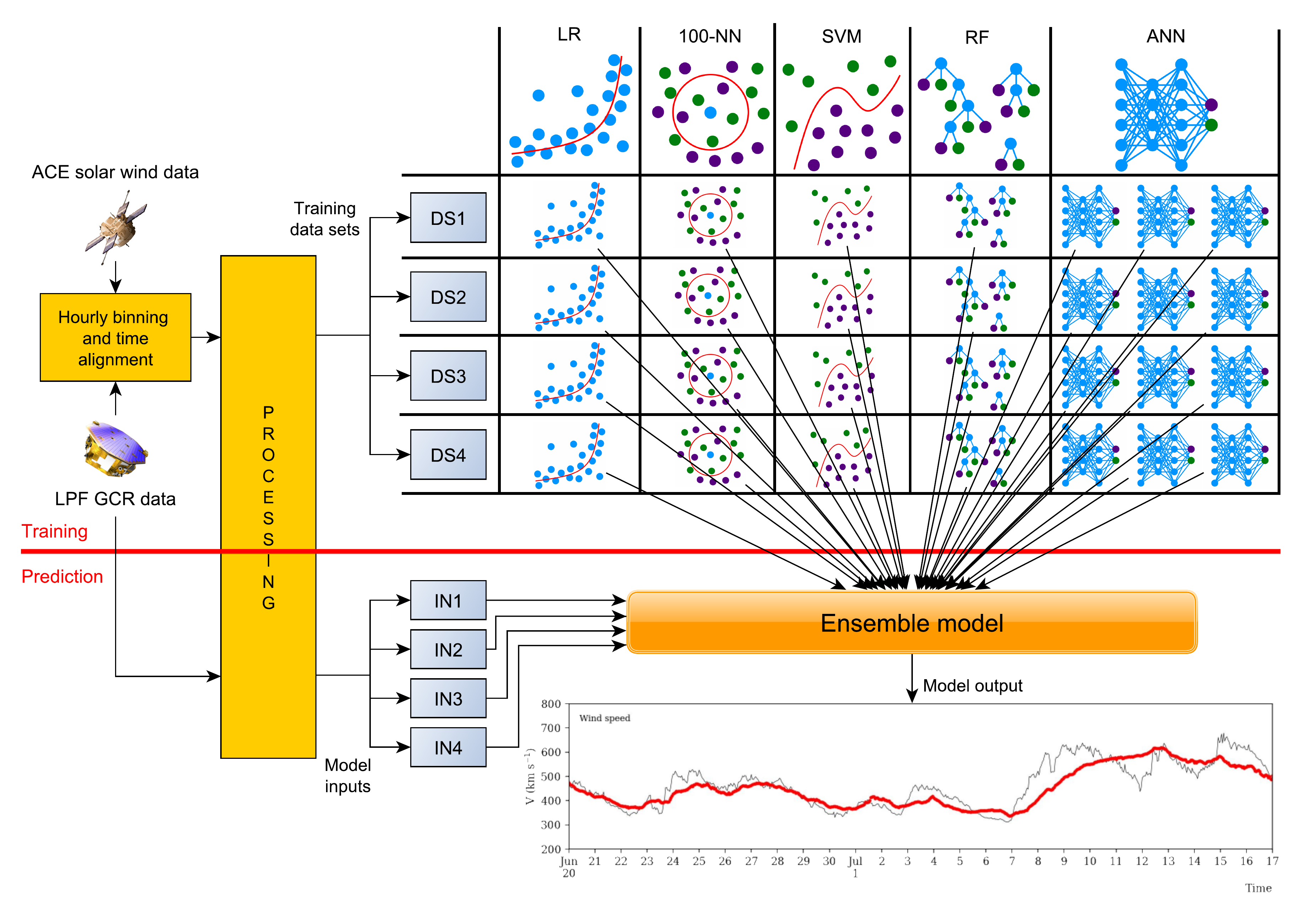}
	\caption{Ensemble model training (top part of the figure) and prediction (bottom part of the figure) workflows. Input, output data and weak predictors are highlighted.}\label{fig:flow}
\end{figure}

The main difficulty in the model training resides in the individual peculiarities of GCR flux short-term variations.
Indeed, as shown in \citet{apj3}, it is possible to observe several GCR flux depressions exhibiting similar characteristics even when associated with different solar wind speed configurations.
On the other hand, similar solar wind speed time profiles are observed during different GCR flux depressions.
This behaviour may depend on interplanetary processes associated with the transit of superposing ICMEs and solar wind HSS.
%
%
We have also observed that similar interplanetary structures generate very different GCR flux depression profiles depending on the passage of different solar wind HSS or ICMEs.
As a matter of fact, intense interplanetary structures modulate differently the overall flux of GCRs depending if maximum values or minimum values of the same are observed at their passage during each BR.

The presented ensemble model relies on the aforementioned base regressors, that during their training phases learn the average relationship binding the trend of GCR flux and the solar wind speed at the passage of high-speed solar wind streams.
As a consequence, when this relationship is influenced by other processes the corresponding data have to be considered as outliers of difficult predictions.
Conversely, the accurate prediction of rare events would highlight an over-fitted model training, resulting in a worse average performance during the model application.

\subsection{The ensemble model}\label{ssec:ens}

A schema of the workflow corresponding to the model training and prediction is reported in \Cref{fig:flow}.
The top part of the figure shows the model training, with required input data, corresponding pre-processing and inner structure.
Conversely, the bottom part is dedicated to the model prediction phase, depicting model inputs, outputs and involved weak predictors.

\begin{table}
	\begin{tabular}{cc@{\hskip 0.4in}cc@{\hskip 0.4in}cc@{\hskip 0.4in}cc@{\hskip 0.4in}cc@{\hskip 0.4in}cc}
		\toprule
		BR & MAE & BR & MAE & BR & MAE & BR & MAE & BR & MAE & BR & MAE \\
		\midrule
		2491 & 55.00 & 2494 & 40.40 & 2497 & 60.13 & 2500 & 67.59 & 2503 & 71.95 & 2506 & 84.98 \\
		2492 & 53.36 & 2495 & 43.89 & 2498 & 75.82 & 2501 & 80.86 & 2504 & 74.57 & 2507 & 66.05 \\
		2493 & 62.35 & 2496 & 56.43 & 2499 & 68.74 & 2502 & 92.01 & 2505 & 82.87 & 2508 & 65.00 \\
		\midrule
		& & & & & & & & & & Mean & 66.78 \\
		\bottomrule
	\end{tabular}
	\caption{Mean absolute error for each BR calculated for the overall ensemble model.}\label{tab:res}
\end{table}

\begin{figure}\centering
	\subfloat[BR 2494.]{
		\includegraphics[width=0.49\linewidth]{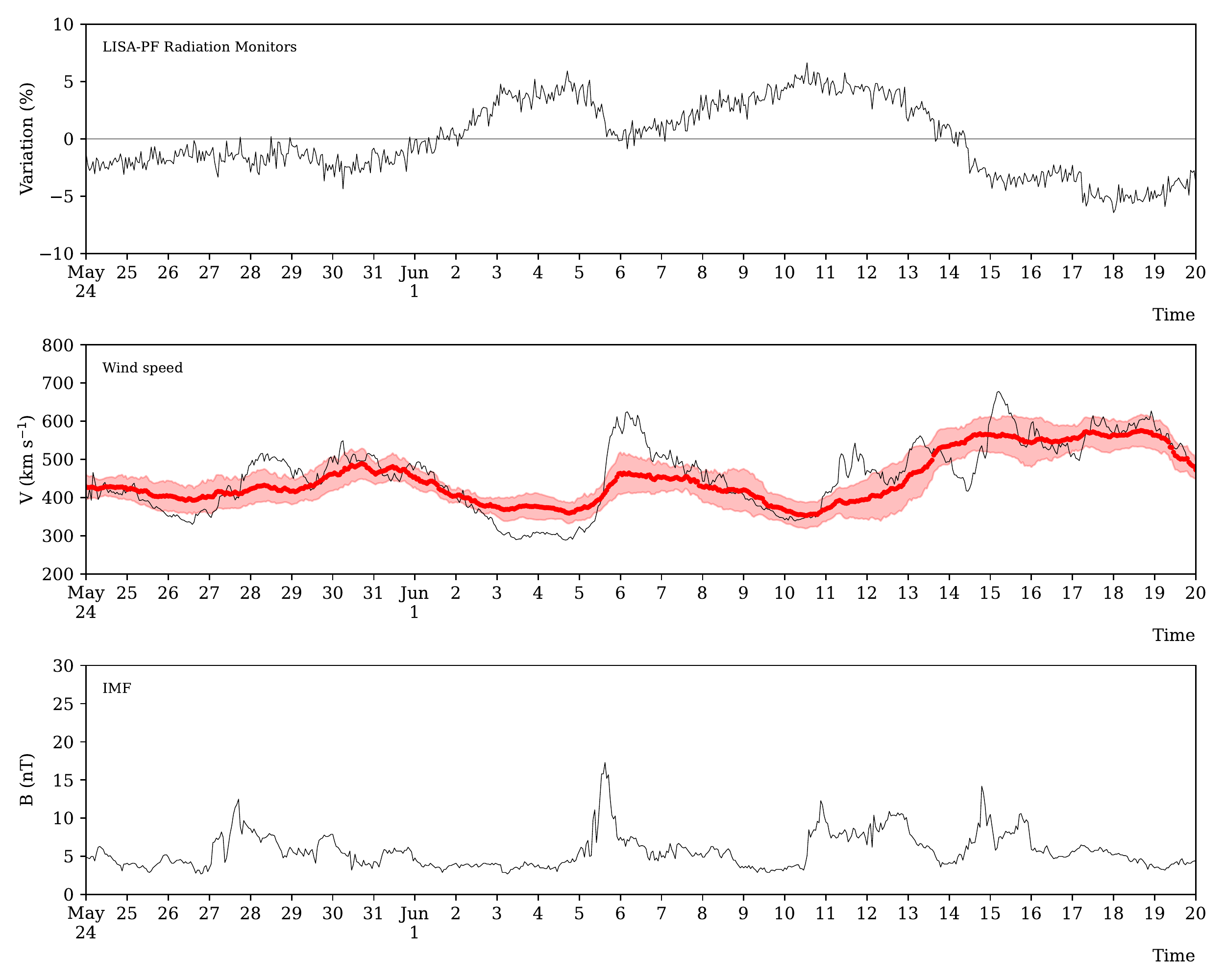}\label{fig:res94}
	}
	\subfloat[BR 2495.]{
		\includegraphics[width=0.49\linewidth]{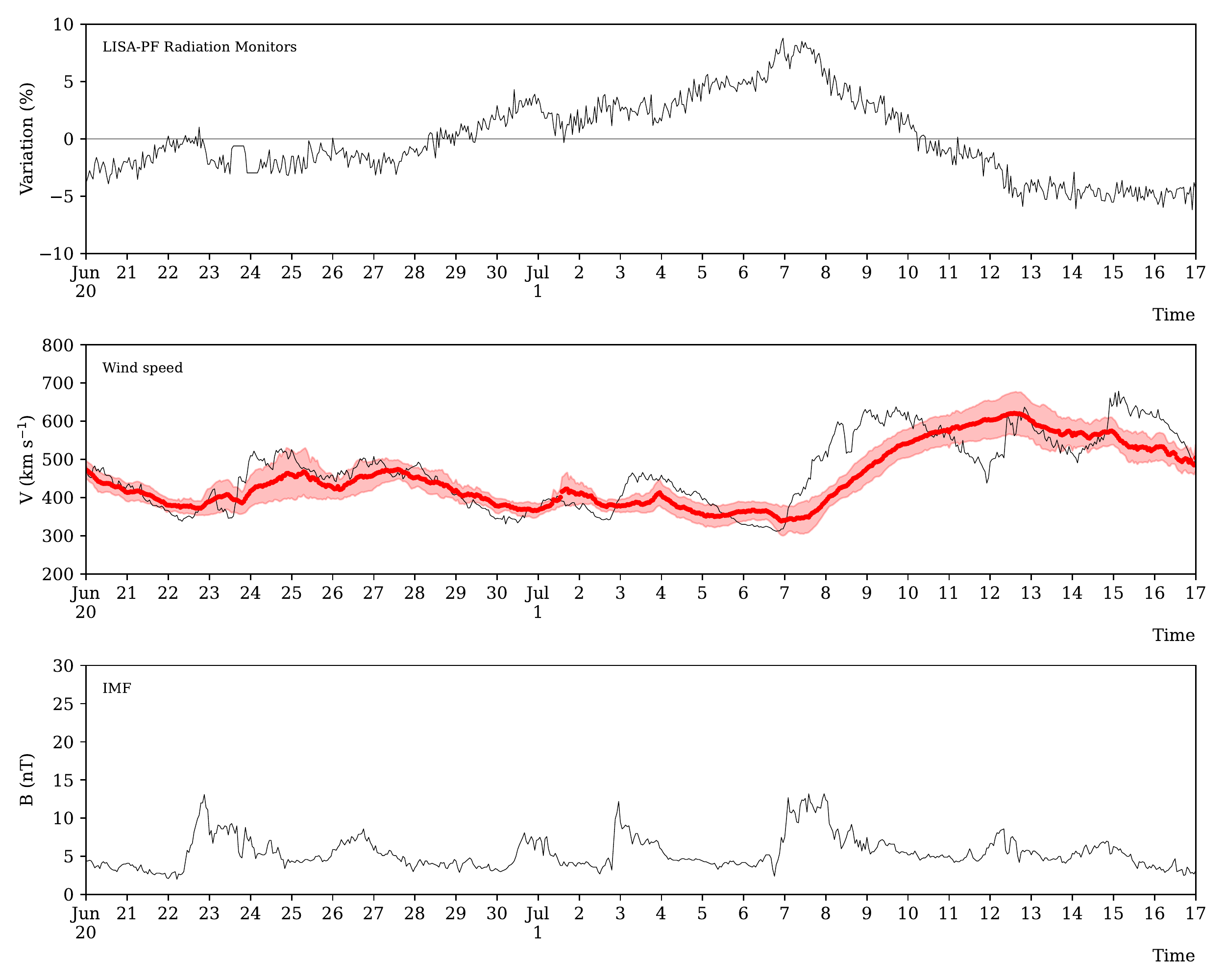}\label{fig:res95}
	}
	\caption{Same as \Cref{fig:lr} for the overall ensemble model. The red shaded region in the middle panels represents the standard deviation calculated on the weak models' predictions.}\label{fig:res}
\end{figure}

The overall ensemble model proposed in this work to carry out a solar wind speed estimate by using as input data GCR flux variations is based on the predictions provided by the described weak regressors.
More in detail, when the model is queried with an input instance to be predicted, 3 steps are executed: \textit{(i)} the instance is processed with the same rationale used for the training data, so 4 different input instances are produced (according to the parameters reported in \Cref{tab:datasets}), as shown in the bottom part of \Cref{fig:flow}; \textit{(ii)} the base models are applied to the processed inputs, resulting in 28 different weak predictions; \textit{(iii)} these predictions are averaged and the resulting value is provided to users as the ensemble model output.

A confidence interval is assigned to the ensemble model outputs, by associating each prediction with the corresponding standard deviation $\sigma$ calculated on all weak models.
As an example, in \Cref{fig:res} the ensemble model predictions for the BRs 2494 and 2495 are reported in the middle panels with red dots, while the confidence interval equal to $\pm \sigma$ around each predicted output is represented as a shaded red region.
We believe that adding a band of uncertainty to the obtained solar wind speed estimates makes them more trustworthy, since also a measure of the weak models' agreement is given to the end user as an indication of prediction reliability.
Indeed, it is possible to observe in the figures that when all the weak models show similar outcomes (and therefore $\sigma$ is small) these last ones are in good agreement with measurements.
Finally, \Cref{tab:res} shows the ensemble regressor MAE for each BR of our data set.
The average predictive error calculated for the overall model is of $\pm$ 70 km s$^{-1}$, corresponding to 10\% of the maximum value of the solar wind speed observed in L1 in 2016--2017 by ACE.

In the future we plan to enhance the performance of our model by providing additional input variables, e.g., IMF parameters.
We also plan to apply symbolic knowledge-extraction techniques in order to obtain human-interpretable clues about the inner behaviour of the presented ML predictor, by considering a similar approach adopted in another work~\citep{sabbatini22LPFSKE}.

\section{Conclusions}\label{sec:conc}

The design of an ML ensemble model allowing us to estimate the solar wind speed trend on the basis of GCR flux observations has been developed for an accurate test-mass charging estimate onboard the LISA mission.
GCR flux variations and solar wind speed measurements were used for the training phase of weak predictors.
The proposed ensemble model predictions allowed for the reconstruction of the solar wind speed observations gathered during 18 BRs in 2016--2017 when the LPF mission remained into orbit around the L1 Lagrangian point.
An average predictive error of $\pm$ 70 km s$^{-1}$ was found w.r.t.\ values of the solar wind speed reaching up to about 700 km s$^{-1}$ during the mission.
The performance of the model may be improved by adding as input data the IMF intensity.
However, the quality of the magnetic field intensity estimates carried out on the basis of data gathered onboard LISA may be limited.
As a result, it was of fundamental importance to optimise our ensemble model with the only benefit of GCR flux measurements.

\section*{Acknowledgements}
The LISA Pathfinder mission is part of the space-science program of the European Space Agency.
The Italian contribution has been supported by the Agenzia Spaziale Italiana and the Istituto Nazionale di Fisica Nucleare.
The LISA Pathfinder data are available at: \url{http://lpf.esac.esa.int/lpfsa/#home}.
Data from the ACE experiments were obtained from the NASA-CDAWeb website: \url{https://cdaweb.gsfc.nasa.gov/}.

\bibliographystyle{apalike}
\bibliography{ensembleV}

\end{document}